\documentclass[11pt]{article}
\usepackage{amssymb}%
\usepackage{amsmath}%

\usepackage{graphicx,amsmath,amsfonts,amscd,amssymb,bm,cite,epsfig,epsf,url,color,algorithm,algorithmic}
\usepackage{fullpage} \usepackage[small,bf]{caption}
\newtheorem{theorem}{Theorem}[section]
\newtheorem{lemma}[theorem]{Lemma}

\newcommand{\E}{\operatorname{\mathbb{E}}}

\newcommand{\cA}{\mathcal{A}}

\newcommand{\bX}{\bm{X}}

\newcommand{\bY}{\bm{Y}}

\newcommand{\be}{\bm{e}}

\newcommand{\tr}{\operatorname{Tr}}

\newcommand{\trace}{\operatorname{Tr}}

\numberwithin{equation}{section}

\def \endprf{\hfill {\vrule height6pt width6pt depth0pt}\medskip}

\renewcommand{\epsilon}{\varepsilon}

\newcommand{\lip}{\text{Lip}}
\newcommand{\Prob}{\mathbb{P}}

\newcommand{\Tp}{T^\perp}

\newcommand{\cS}{\mathcal{S}}

\title{Quantum tomography from few full-rank observables.}

\author{ Vladislav Voroninski\\
  \vspace{-.1cm}\\
  Department of Mathematics, Massachusetts Institute of Technology. 
}

\date{September 2013}

\begin{document}
\maketitle

\section{Introduction}

It was proven in \cite{CSV} that the PhaseLift algorithm recovers signals $x \in \mathbb{C}^n$ exactly from $m = O(n \log n)$ measurements $\{\left| \left< x, z_i \right> \right|^2\}_{i=1}^m$ with high probability when the measurement vectors $z_i \in \mathbb{C}^n$ are iid gaussian and that this procedure is provably stable with respect to measurement noise under the same assumptions. To be precise, this means that in the noiseless case, for a fixed $x \in \mathbb{C}^n$ and defining the linear operator $\cA: X \in \mathbb{C}^{n \times n} \mapsto \{\tr(X z_i z_i^*) \}_{i=1}^m$, the program
\begin{equation}
\label{eq:tracemin}
 \begin{array}{ll}
    \text{minimize}   & \quad \trace(X)\\ 
    \text{subject to} & \quad  \cA(X) = \cA(xx^*)\\
& \quad X \succeq 0;  
\end{array}
\end{equation}
recovers $xx^*$ with high probability. The stability result uses a modified, noise-aware convex program. These guarantees were subsequently improved to hold uniformly over all signals for $m = O(n)$ with sharp stability guarantees in \cite{Five} and it was shown in \cite{SOR} that in the noiseless case, this program has only one point in its feasible set, namely $xx^*$. 

However, the gaussian measurement model is not know to be physically realizable. Therefore it is of interest to prove exactness results for PhaseLift under more structured measurement assumptions, which requires more technical proofs due to the lack of probabilistic independence between sensing vectors in structured random measurement ensembles. 

A step in this direction is to consider the $z_i$ as rows of iid Haar distributed unitary matrices, a situation that occurs in quantum tomography from measurements with full-rank observables. In this paper, we prove that PhaseLift succeeds with high probability under this measurement model as long as the number of observables is $O(1)$ (which corresponds to $m$ = O(n) in the above setting) and point out a corollary of the result which relates to Wright's conjecture. This conjecture, that there exist 3 observables which determine any pure state, has been recently disproven in \cite{HMW11, DMBlog} and a close variant of it, that 4 generic observables suffice to determine any pure state, was recently settled in \cite{DAP}. This brings us to the main theorem:

\begin{theorem}
Take $x \in \mathbb{C}^n$ and assume that measurements of the form $\{|U_{k}x|^2\}_{k=1}^r$ are available, where the $U_i$ are sampled independently according to the Haar measure on $\mathbb{U}(n)$, the unitary group or $\mathbb{O}(n)$, the orthogonal group, so that the total number of measurements is $m = rn$. Then the PhaseLift algorithm succeeds in recovering $x$ up to global phase with very high probability with $m = \text{O}(n)$. 
\end{theorem}
Here we assume that measurements of the form $\{|U_{k}x|^2\}_{k=1}^r$ are available, where the $U_i$ are sampled independently according to the Haar measure on $\mathbb{U}(n)$, the unitary group or $\mathbb{O}(n)$, the orthogonal group, and the total number of measurements is $m = rn$. Below, we will label the transpose of the row vectors of $U_{k}$ as $u_{i}^{(k)}$ or enumerate them as $\{u_{i}\}_{i=1}^{m}$. As in the gaussian case, we may assume wlog that $x = e_1$, in this case by the unitary/orthogonal invariance of the Haar measure. 

We proceed by showing that the measurement operator $\cA$ in this setting obeys some nice properties with high probability. Namely, we need to verify that $\cA$ satisfies the condition of the following lemma, which is a very slight modification of Lemma 3.6.4 in \cite{CSV} achieved by noting that if $Y_\Tp \prec 0$, then $\left<H_\Tp,Y_\Tp \right> \leq 0$. 
\begin{lemma}
  \label{lem:crucialcomplex2}
  Suppose that the mapping $\cA$ obeys the following two properties:
  for some $\delta \le 3/13$: 
  
  1) for all positive semidefinite matrices $\bX$,
\begin{equation}
\label{eq:RIP1r1C}
m^{-1} \|\cA(\bX)\|_1 \leq (1+\delta) \|\bX\|_{1};
\end{equation}
2) for all matrices $\bX \in T$
\begin{equation}
\label{eq:RIP1r2C}
m^{-1} \|\cA(\bX)\|_1 \geq 2(\sqrt{2}-1)(1-\delta)\|\bX\| \geq
0.828(1-\delta) \|\bX\|.
\end{equation}
Suppose further that there exists $Y$ in the range of $\cA^*$ obeying
\begin{equation}
  \label{eq:dualcertifC}
  \|\bY_T - \be_1 \be_1^*\|_2 \le 1/5 \quad \text{and} \quad Y_\Tp \prec 0.
\end{equation}
Then $\be_1 \be_1^*$ is the unique minimizer of PhaseLift.
\end{lemma}

In particular, the RIP-1 property in this unitary case has implications related to Wright's conjecture. Furthermore, we adapt a trick in the construction of the dual certificate, used by \cite{Five} in the gaussian case, to reduce the number of necessary measurements from $O(n\log n)$ to $O(n)$. Establishing the above yields the main result.

\section{Restricted Isometry Property of type 1 for Unitary Matrices}
 In the sequel we will label the transpose of the row vectors of $U_{k}$ as $u_{i}^{(k)}$ or enumerate them as $\{u_{i}\}_{i=1}^{m}$. As in the gaussian case, we may assume wlog that $x = e_1$, in this case by the unitary/orthogonal invariance of the Haar measure.\\

First, we aim to establish a RIP-1 property on rank-2 matrices  for this class of measurements. Let $\cA(X) = \{\sqrt{n(n+1)} \tr(u_{i} u_{i}^* X) \}_{i=1}^m$, where $\cA$ is a linear map from the Hermitian matrices. Let $X = x_1 x_1^* - \lambda x_2 x_2^*$ be a rank-2 hermitian matrix in SVD form with $0\leq \lambda \leq 1$. Then
\begin{align*}
\frac{1}{\sqrt{n(n+1)}}\cA(x_1 x_1^* - \lambda x_2 x_2^*) &= \{ |\left<u_i,x_1 \right>|^2 - \lambda |\left<u_i,x_2\right>|^2 \}_{i=1}^{m}  \\
&=^d  \{ |\left<u_i,e_1 \right>|^2 - \lambda |\left<u_i,e_2\right>|^2 \}_{i=1}^{m}\\
& = \{ |u_{i1}|^2 - \lambda |u_{i2}|^2 \}_{i=1}^{m}
\end{align*} 
where we used rotational invariance of Haar measure and the fact that there exist orthogonal or unitary transformations taking any real/complex orthobasis to another orthobasis and $u_{ij}$ denotes the $j$th entry  of the vector $u_i$. 

To establish $\frac{1}{m}\|\cA(X)\|_1 \leq (1+\delta) \|X\|_1$ for all psd matrices, it is enough to consider $X$ to be rank 1 psd. Taking any unit vector $x \in \mathbb{C}^n$, we have
\[
\frac{1}{r}\frac{1}{\sqrt{n(n+1)}}\|\cA(xx^*)\|_{l1} = \frac{1}{r}\sum_{i=1}^m |\left<u_{m},x\right>|^2 = 1
\]
This implies that 
\[
\frac{1}{r}\frac{1}{\sqrt{n(n+1)}}\|\cA(X)\|_1 \leq (1+\delta) \|X\|_1
\]
for any $\delta > 0$ and for any psd $X$. Now since $\frac{m}{r\sqrt{n(n+1)}} = \sqrt{\frac{n}{n+1}}$, we can get the desired property with $\delta = \frac{3}{13}$.  

To get the other part of RIP-1, we need to examine the quantity
\[
\frac{1}{r}\frac{1}{\sqrt{n(n+1)}}\|\cA(x_1 x_1^* - \lambda x_2 x_2^*)\|_{l1}  =^d \frac{1}{r}\sum_{i=1}^{m} | |u_{i1}|^2 - \lambda |u_{i2}|^2 | = \frac{1}{r}\sum_{k=1}^{r}\sum_{i=1}^{n} | |u_{i1}^{(k)}|^2 - \lambda |u_{i2}^{(k)}|^2 | 
\]
and show that it is lower bounded by a multiple of the operator norm of $X = x_1 x_1^* - \lambda x_2 x_2^*$ whp. This sum may be expressed as a function of $2rn$ iid gaussian rvs. This function is not Lipschitz, so we will use a surrogate function that is Lipschitz in order to apply Talagrand's inequality \cite{Two} and then show that this introduces only a very small error. 

We will treat the real and complex cases simultaneously. To be specific, one way to obtain the Haar measure on $\mathbb{O}(n)$ or $\mathbb{U}(n)$ is to perform Gram-Schmidt on the columns of a gaussian or complex gaussian matrix. Thus, we will consider the columns $u_1$ and $u_2$ of a Haar-distributed orthogonal matrix as the result of the Gram-Schmidt procedure on a pair of iid gaussian vectors $\zeta$ and $z$. Introduce the functions $v(x) = \frac{x}{\|x\|_2}$ and $t(x,y) = x - y\left<y,x\right>$. Then if $\zeta, z$ are iid $\mathcal{N}(0,I)$ or $\mathcal{C}\mathcal{N}(0,I,0)$, it can be verified that 
\[
(u_1,u_2) =^d (v(z),v(t(v(\zeta),v(z))))
\]
We can now express the distribution of the quantity above as 
\begin{align*}
\frac{1}{r}\sum_{i=1}^{m} | |u_{i1}|^2 - \lambda |u_{i2}|^2 |   &=^d \frac{1}{r}\sum_{k=1}^r F(\zeta^{(k)},z^{(k)})\\
&= \frac{1}{r}\sum_{k=1}^{r}\sum_{i=1}^{n} | |v(z^{(k)})_i|^2 - \lambda |v(t(v(\zeta^{(k)}),v(z^{(k)})))_i|^2 | 
\end{align*}

as a function of a $2rn$ component gaussian vector. The above function is not lipschitz and the issue occurs in two places: first, when we normalize the vectors $\zeta$ and $z$ and then when we normalize the expression $t(v(\zeta),v(z))$. Let us introduce the surrogate functions $\tilde{v}_{1}(x) = \frac{x}{(\|x\|_2 \vee \sqrt{\frac{n}{c_1}})}$ where $c_1 >>1$ and $\tilde{v}_2 = \frac{x}{(\|x\|_2 \vee \sqrt{c_2})}$ where $0< c_2< 1$. Now, the surrogate function
\[
\frac{1}{r}\sum_{k=1}^r \tilde{F}(\zeta^{(k)},z^{(k)})\\ = \frac{1}{r}\sum_{k=1}^{r}\sum_{i=1}^{n} | |\tilde{v}_1 (z^{(k)})_i|^2 - \lambda |\tilde{v}_2 (t(\tilde{v}_1 (\zeta^{(k)}),\tilde{v}_1 (z^{(k)})))_i|^2 | 
\]
is equal to the original with probability at least 
\[
 1 - 2r\mathbb{P}\{ \|\zeta\|_2^2 < \frac{n}{c_1} \} - r\mathbb{P}\{ |\left<\frac{\zeta}{\|\zeta\|_2},\frac{z}{\|z\|_2} \right>|^2 > 1- c_2\} \geq 1- \text{O}\left(re^{-\gamma n}\right)
 \]
 where $\gamma$ can be made arbitrarily large by taking $c_1$ large enough and $c_2$ small enough. It now remains to verify that the surrogate function is Lipschitz with a good enough constant and that it introduces only a small error.\\

Consider the function $g(x,y) = \sum_{i=1}^{n}||x_i|^2 - \lambda |y_i|^2|$, with $x,y \in \mathbb{R}^n$ or $\mathbb{C}^n$ and assume $\|x_i\|_2 \leq 1, \|y_i\|_2 \leq 1$. We have
\begin{align*}
|g(x_1,y_1) - g(x_2,y_2)| &\leq \sum_{i=1}^{n} \left| ||x_{1i}|^2 - \lambda |y_{1i}|^2| - ||x_{2i}|^2 - \lambda |y_{2i}|^2| \right|\\
&\leq \sum_{i=1}^{n} |(|x_{1i}|^2 - |x_{2i}|^2) - \lambda (|y_{1i}|^2-|y_{2i}|^2)|\\
&\leq \sum_{i=1}^{n} |(|x_{1i}| + |x_{2i}|)(|x_{1i}| - |x_{2i}|)| + \lambda |(|y_{1i}|+|y_{2i}|)(|y_{1i}|-|y_{2i}|)|\\
&\leq \||x_1|+|x_2|\|_2 \|x_1 - x_2 \|_2 + \lambda \||y_1| +| y_2|\|_2 \|y_1 - y_2 \|_2\\
& \leq 2\|x_1 - x_2 \|_2 + 2\lambda\|y_1 - y_2 \|_2 
\end{align*}

Now take
\begin{align*}
&|\tilde{F}(\zeta_1,z_1) - \tilde{F}(\zeta_2,z_2)| \\
&= |g(\tilde{v}_1 (z_1),\tilde{v}_2 (t(\tilde{v}_1 (\zeta_1),\tilde{v}_1 (z_1)))) - g(\tilde{v}_1 (z_2),\tilde{v}_2 (t(\tilde{v}_1 (\zeta_2),\tilde{v}_1 (z_2))))|\\
&\leq 2\| \tilde{v}_1 (z_1) - \tilde{v}_1 (z_2)  \|_2 + 2\lambda\|   \tilde{v}_2 (t(\tilde{v}_1 (\zeta_1),\tilde{v}_1 (z_1))) - \tilde{v}_2 (t(\tilde{v}_1 (\zeta_2),\tilde{v}_1 (z_2))) \|_2 \\
&\leq 2\lip(\tilde{v}_1)\|z_1-z_2\|_2 + 2\lambda \lip(\tilde{v}_2) \| t(\tilde{v}_1 (\zeta_1),\tilde{v}_1 (z_1)) - t(\tilde{v}_1 (\zeta_2),\tilde{v}_1 (z_2)) \|_2 \\
&\leq 2\lip(\tilde{v}_1)\|z_1-z_2\|_2 + 2\lambda \lip(\tilde{v}_2) \lip(t\vert_{\mathcal{B}(0,1)^2}) \| (\tilde{v}_1 (\zeta_1),\tilde{v}_1 (z_1)) - (\tilde{v}_1 (\zeta_2),\tilde{v}_1 (z_2)) \|_2\\
& \leq 2\lip(\tilde{v}_1)\|z_1-z_2\|_2 + 2\lambda \lip(\tilde{v}_2) \lip(t\vert_{\mathcal{B}(0,1)^2}) \lip(\tilde{v}_1) \| ( \zeta_1,z_1) - (\zeta_2, z_2) \|_2\\
\end{align*}

One can verify that in either the real or complex case, when $\|x_i\|_2,\|y_i\|_2 \leq 1$, the function $t$ satisfies
\[
\|t(x_1,y_1) - t(x_2,y_2)\|_2 \leq 2\| (x_1,y_1)-(x_2,y_2)  \|_2
\]

%

For $x \in \mathbb{R}^n$, let $\tilde{v}(x) = \frac{x}{\|x\|_2 \vee c}$ for some positive constant c. Now, we have that $D(\frac{x}{\|x\|_2}) = \frac{1}{\|x\|_2^3}(\|x\|_2^2I - xx^*)$ and hence $\|D(\frac{x}{\|x\|_2})\| \leq \frac{2}{\|x\|_2}$. Thus, $\|D(\tilde{v})\| \leq \frac{2}{c}$ on $\bar{\mathcal{B}}(0,c)^{c}$ so that $\lip(\tilde{v}|_{U}) \leq \frac{2}{c}$ for any open convex set $U \in \bar{\mathcal{B}}(0,c)^{c}$ . Furthermore, we have $D(\tilde{v}) = \frac{1}{c}I$ on $\mathcal{B}(0,c)$ and thus $\|D(\tilde{v})\| \leq \frac{1}{c}$ on $\mathcal{B}(0,c)$ so that $\lip(\tilde{v}|_{\bar{\mathcal{B}}(0,c)}) \leq \frac{1}{c}$.\\

Take $x_i \in \bar{\mathcal{B}}(0,c)^c, i = 1,2$ such that the line connecting these two points intersects $\bar{\mathcal{B}}(0,c)$. Assume that the point(s) of intersection are $z_1$ and $z_2$ (with the line from $x_1$ to $x_2$ first hitting $z_1$ and then $z_2$).  Then we have 

\begin{align*}
\|\tilde{v}(x_1) - \tilde{v}(x_2)\|_2 &\leq \|\tilde{v}(x_1) - \tilde{v}(z_1)\|_2+\|\tilde{v}(z_1) - \tilde{v}(z_2)\|_2+\|\tilde{v}(z_2) - \tilde{v}(x_2)\|_2\\
& \leq \frac{2}{c} \|x_1 - z_2\|_2 + \frac{1}{c}\|z_1-z_2\|_2 + \frac{2}{c}\|z_2 - x_2 \|_2\\
& \leq \frac{2}{c}\|x_1-x_2\|_2
\end{align*}

where for the sets $U_i$ we take $\mathcal{B}(x_i,\|z_i-x_i\|_2)$. The other cases of arrangements of $x_i$ are similarly proven. We conclude that $\lip(\tilde{v}) \leq \frac{2}{c}$. Note that this implies that in the complex case, we also have $\lip(\tilde{v}) \leq \frac{2}{c}$ with $\tilde{v}$ defined analogously. We have thus established that  $\lip(\tilde{v}_1) \leq \frac{\sqrt{c_2}}{\sqrt{n}}$ and $\lip(\tilde{v}_2) \leq \frac{2}{\sqrt{c_1}}$. Using this information, 
\[
|\tilde{F}(\zeta_1,z_1) - \tilde{F}(\zeta_2,z_2)| \leq \frac{16\sqrt{\frac{c_1}{c_2}}}{\sqrt{n}}\|(\zeta_1,z_1)-(\zeta_2,z_2)\|_2
\]
Finally, this implies 
\[
\lip(\frac{1}{r}\sum_{k=1}^r \tilde{F}(\zeta^{(k)},z^{(k)}) )\leq \frac{1}{\sqrt{r}}\frac{16\sqrt{\frac{c_1}{c_2}}}{\sqrt{n}}
\]
By Talagrand's inequality, we have
\[
\mathbb{P}\{ \left|\frac{1}{r}\sum_{k=1}^r \tilde{F}(\zeta^{(k)},z^{(k)}) - \E\left[\frac{1}{r}\sum_{k=1}^r \tilde{F}(\zeta^{(k)},z^{(k)})  \right]\right| \geq t \} \leq e^{-c(rn)t^2}
\]
for a constant c which depends on $c_i$. Let 
\[
\tilde{G} = \frac{1}{r}\sum_{k=1}^r \tilde{F}(\zeta^{(k)},z^{(k)}), \qquad G = \frac{1}{r}\sum_{k=1}^r F(\zeta^{(k)},z^{(k)}).
 \]
 $G$ and $\tilde{G}$ are both bounded by 2 and disagree on a set of probability $\text{O}\left(re^{-\gamma n}\right)$, thus 
 \[
 \lim_{n \rightarrow \infty} \left| \E[\tilde{G}] - \E\left[G\right] \right|=0.
 \]
 So that if we fix $t$ apriori, then for all $n$ large enough
\begin{align*}
&\mathbb{P}\{ \left|G -\E\left[G \right]\right| \geq t  \}\\
&\leq \mathbb{P}\{ \left|\tilde{G} - \E[\tilde{G} ] \right| \geq t  - \left|\tilde{G}-G \right| - \left| \E[\tilde{G}] - \E\left[G\right] \right| \}\\
&\leq \mathbb{P}\{ G \neq \tilde{G} \} + \mathbb{P}\{ \left|\tilde{G} - \E[\tilde{G}] \right| \geq t  - \left| \E[\tilde{G}] - \E\left[G\right] \right| \}\\
&\leq  \text{O}\left(re^{-\gamma n}\right) + e^{-c(rn)(t/2)^2}
\end{align*}

Therefore, we have established
\[
\mathbb{P}\{ \left| \frac{1}{r}\sum_{i=1}^{m} | |u_{i1}|^2 - \lambda |u_{i2}|^2 |  - \E\left[\frac{1}{r}\sum_{i=1}^{m} | |u_{i1}|^2 - \lambda |u_{i2}|^2 |   \right]\right| \geq t \} \leq e^{-c(rn)(t/2)^2} + \text{O}\left(re^{-\gamma n}\right)
\]
for constants c and $\gamma$ which depend on $c_i$. To achieve an arbitrarily fast exponential rate, first select $c_i$ so that $\gamma$ is as large as needed, then fix r large enough.\\
 
 We claim that $\E\left[ \left| \left|u_{i1}\right|^2 - \lambda |u_{i2}|^2  \right| \right] = \frac{1}{n}\frac{1+\lambda^2}{1+\lambda}$, which we compute below. We have from $\left[4 \right]$ that $\left(|u_{i1}|^2, \ldots, |u_{i n-1}|^2 \right)$ are uniformly distributed on $\{(x_1,\ldots x_{n-1}); x_i \geq 0, \sum_{i=1}^{n-1}x_i \leq 1\}$. Thus
 \begin{align*}
& \E\left[ \left| \left|u_{i1}\right|^2 - \lambda |u_{i2}|^2  \right| \right] \frac{1}{(n-1)(n-2)}\\
&= (n-3)!\int_{\mathbb{R}^{n-1}}|x_1-\lambda x_2| \chi_{\{\sum_{i=1}^{n-1}x_i \leq 1, x_i \geq 0\}} dx_1 \ldots dx_{n-1}\\
 &= (n-3)!\int_{\mathbb{R}^2} |x_1-\lambda x_2| \chi_{\{x_1 + x_2 \leq 1, x_i \geq 0\}} \int_{\mathbb{R}^{n-3}} \chi_{\{x_3 + \ldots x_{n-1} \leq 1- (x_1 + x_2)\}} dx_3 \ldots dx_{n-1}\\
 &= \frac{(n-3)!}{(n-3)!}\int_{\mathbb{R}^2}|x_1 - \lambda x_2| (1-(x_1+x_2))^{n-3} \chi_{\{x_1+x_2 \leq 1, x_i \geq 0\}}dx_1 dx_2\\
 &=\int_{0}^{1} \int \left[ \chi_{\{x_1 \leq \lambda x_2\}}(\lambda x_2 - x_1) + \chi_{\{x_1 \geq \lambda x_2\}}(x_1-\lambda x_2)\right](1-(x_1+ x_2))^{n-3}\chi_{\{0\leq x_2 \leq 1-x_2\}} dx_1 dx_2\\
 &=\int_{0}^1 \chi_{\{\lambda x_2 \leq 1-x_2\}} \int_{0}^{\lambda x_2}(\lambda x_2 - x_1)(1-(x_1+x_2))^{n-3}dx_1 \\
 &+ \chi_{\{\lambda x_2 \geq 1-x_2\}} \int_{0}^{1-x_2} (\lambda x_2 - x_1)(1-(x_1+x_2))^{n-3} dx_1 \\
 &  +     \chi_{\{\lambda x_2 \leq 1-x_2\}} \int_{\lambda x_2}^{1-x_2} (\lambda x_2 - x_1)(1-(x_1+x_2))^{n-3} dx_1 dx_2\\
 & = \int_{0}^{\frac{1}{1+\lambda}} \int_{0}^{\lambda x_2} (\lambda x_2 - x_1)(1-(x_1+x_2))^{n-3}  dx_1 + \int_{\lambda x_2}^{1-x_2} ( x_1 - \lambda x_1)(1-(x_1+x_2))^{n-3} dx_1 dx_2 \\
 & +\int_{\frac{1}{1+\lambda}}^{1} \int_{0}^{1-x_2} (\lambda x_2 - x_1)(1-(x_1+x_2))^{n-3}  dx_2 dx_2\\
 \end{align*}
 \begin{align*}
 &= \int_{0}^{\frac{1}{1+\lambda}}  \lambda x_2\left( \frac{-1}{n-2}(1-(x_1+x_2))^{n-2} \middle\vert_{0}^{\lambda x_2} \right) - \int_{0}^{\lambda x_2} x_1(1-(x_1+x_2))^{n-3} dx_1  dx_2 \\
 &+ \int_{0}^{\frac{1}{1+\lambda}}  \int_{\lambda x_2}^{1-x_2} x_1(1-(x_1+x_2))^{n-3} dx_1 - \lambda x_2 \left( \frac{-1}{n-2}(1-(x_1+x_2))^{n-2} \middle \vert_{\lambda x_2}^{1-x_2} \right) dx_2\\
 &+ \int_{\frac{1}{1+\lambda}}^{1} \lambda x_2\left( \frac{-1}{n-2}(1-(x_1+x_2))^{n-2} \middle\vert_{0}^{1-x_2} \right) - \int_{0}^{1-x_2}x_1(1-(x_1+x_2))^{n-3}dx_1 dx_2\\
 &= \int_{0}^{\frac{1}{1+\lambda}}  \lambda x_2\left( \frac{-1}{n-2}(1-(1+\lambda)x_2))^{n-2} +\frac{1}{n-2}(1-x_2)^{n-2} \right) \\
 &- \int_{0}^{\lambda x_2} x_1(1-(x_1+x_2))^{n-3} dx_1  dx_2 \\
 &+ \int_{0}^{\frac{1}{1+\lambda}}  \int_{\lambda x_2}^{1-x_2} x_1(1-(x_1+x_2))^{n-3} dx_1 - \lambda x_2 \left( \frac{1}{n-2}(1-(1+\lambda)x_2))^{n-2} \right) dx_2\\
&+ \int_{\frac{1}{1+\lambda}}^1 \lambda x_2\left( \frac{1}{n-2} (1-x_2)^{n-2}  \right)dx_1 - \int_{0}^{1-x_2} x_1 (1-(x_1+x_2))^{n-3}dx_1 dx_2\\
&= \frac{\lambda}{(1+\lambda)^2} \frac{-1}{n-2}\int_{0}^{1} x_2(1-x_2)^{n-2} dx_2 + \lambda \frac{1}{n-2} \int_{0}^{\frac{1}{1+\lambda}} x_2(1-x_2)^{n-2}dx_2\\
&- \int_{0}^{\frac{1}{1+\lambda}} \frac{-1}{n-2}x_1(1-(x_1+x_2))^{n-2}\left\vert_{0}^{\lambda x_2}  -\frac{1}{(n-2)(n-1)}(1-(x_1+x_2))^{n-1} \right\vert_{0}^{\lambda x_2}  dx_2\\
&+\int_{0}^{\frac{1}{1+\lambda}} \frac{-1}{n-2}x_1(1-(x_1+x_2))^{n-2} \left\vert_{\lambda x_2}^{1-x_2} - \frac{1}{(n-2)(n-1)}(1-(x_1+x_2))^{n-1} \right\vert_{\lambda x_2}^{1-x_2}\\
& \frac{-\lambda}{(1+\lambda)^2}\frac{1}{n-2}\int_{0}^{1}x_2(1-x_2)^{n-2}dx_2 + \frac{\lambda}{n-2} \int_{\frac{1}{1+\lambda}}^{1} x_2(1-x_2)^{n-2} dx_2\\
&- \int_{\frac{1}{1+\lambda}}^{1} \frac{-1}{n-2}x_1(1-(x_1+x_2))^{n-2} \left\vert_{0}^{1-x_2} - \frac{1}{(n-2)(n-1)}(1-(x_1+x_2))^{n-1}\right\vert_{0}^{1-x_2} dx_2\\
\end{align*}
\begin{align*}
&=\frac{\lambda}{(1+\lambda)^2}\frac{-1}{n-2}\int_{0}^1 x_2(1-x_2)^{n-2}dx_2 + \frac{\lambda}{n-2}\int_{0}^{\frac{1}{1+\lambda}}x_2(1-x)2)^{n-2}dx_2\\
&-\int_{0}^{\frac{1}{1+\lambda}} \frac{-1}{n-2}\lambda x_2(1-(1+\lambda)x_2)^{n-2} -\frac{1}{(n-2)(n-1)}(1-(1+\lambda)x_2)^{n-1} \\
&+ \frac{1}{(n-2)(n-1)}(1-x_2)^{n-1} dx_2 + \int_{0}^{\frac{1}{1+\lambda}} \frac{1}{n-2}\lambda x_2(1-(1+\lambda)x_2)^{n-2}\\
& + \frac{1}{(n-2)(n-1)}(1-(1+\lambda)x_2)^{n-1} dx_2\\
&+ \frac{-\lambda}{(1+\lambda)^2}\frac{1}{n-2} \int_{0}^{1} x_2(1-x_2)^{n-2} dx_2 + \frac{\lambda}{n-2}\int_{\frac{1}{1+\lambda}}^1 x_2(1-x_2)^{n-2} dx_2\\
&-\int_{\frac{1}{1+\lambda}}^1 \frac{1}{(n-2)(n-1)}(1-x_2)^{n-1} dx_2\\
&= \frac{\lambda}{(1+\lambda)^2)}\frac{-1}{n-2}\int_{0}^{1} x(1-x)^{n-2} dx + \frac{\lambda}{n-2}\int_{0}^{\frac{1}{1+\lambda}}x(1-x)^{n-2}dx\\
&+\frac{\lambda}{(1+\lambda)^2}\frac{1}{n-2}\int_{0}^{1} x(1-x)^{n-2} dx + \frac{1}{1+\lambda}\frac{1}{(n-2)(n-2)}\int_{0}^{1}(1-x)^{n-1}dx\\
&-\frac{1}{(n-2)(n-1)}\int_{0}^{\frac{1}{1+\lambda}}(1-x)^{n-1} dx + \frac{\lambda}{(1+\lambda)^2}\frac{1}{n-2}\int_{0}^{1} x(1-x)^{n-2} dx\\
&+ \frac{1}{1+\lambda}\frac{1}{(n-2)(n-2)}\int_{0}^{1}(1-x)^{n-1}dx -\frac{\lambda}{(1+\lambda)^2}\frac{1}{n-2}\int_{0}^{1} x(1-x)^{n-2} dx\\
&+\frac{\lambda}{n-2}\int_{\frac{1}{1+\lambda}}^1 x(1-x)^{n-2}dx - \frac{1}{(n-2)(n-1)}\int_{\frac{1}{1+\lambda}}^1 (1-x)^{n-1}dx\\
&= \frac{\lambda}{n-2}\int_{0}^1 x(1-x)^{n-2} dx + \left[ 2\frac{1}{1+\lambda} - 1 \right]\frac{1}{(n-2)(n-1)} \int_{0}^1 (1-x)^{n-1} dx \\
&=\frac{1}{n(n-1)(n-2)}\left[ \lambda + \frac{1-\lambda}{1+\lambda} \right]\\
&=\frac{1}{n(n-1)(n-2)}\frac{1+\lambda^2}{1+\lambda}
 \end{align*}
 Thus
 \[
\E\left[\frac{1}{r}\sum_{i=1}^{m} | |u_{i1}|^2 - \lambda |u_{i2}|^2 |   \right] = \frac{1+\lambda^2}{1+\lambda}
 \]
 which, as in the complex gaussian case, achieves its minimum on $\left[ 0,1\right]$ of $2(\sqrt{2}-1)>0.828$.\\

 \subsection{Implications related to Wright's conjecture}
 Using the same covering argument over rank-2 indefinite matrices as in Lemma 3.4.2 in \cite{CSV}, we obtain the RIP-1 property for unitary matrices. Since RIP-1 is stronger than injectivity of the measurements, this shows that there exists some integer $r$ such that the measurements $|U_i x|_{i=1}^r$, where $U_i$ are iid Haar distributed unitary matrices, are injective up to global phase with very high probability. It would be interesting to see how small of an integer $r$ can be achieved by probabilistic arguments, say by using more sophisticated concentration arguments, but there seems to be a bottleneck in the large constants that appear in concentration inequalities. On the other hand, algebraic and differential geometry techniques are successful in establishing that an RIP-1 property with some nonzero, possibly very small constant holds for 4 unitary matrices \cite{DAP}.
 
 \section{Dual certification}
 We start with a useful property:
\subsection{Moments of entries of a unitary matrix}
Wlog, we shall further treat below the complex case only. We record some useful identities from \cite{One}. Let $u_{ij}$ be an entry of a $n \times n$ Haar distributed unitary matrix. Then
\[
\mathbb{E}[ |u_{ij}|^{2d} ] = \frac{d!}{n(n+1)\ldots(n+d-1)} 
\]
Which implies that $ \E[|u_{ia}|^4] = \frac{2}{n(n+1)}$.
Using the identity 
\[
\frac{1}{n} = \mathbb{E}[|u_{ia}|^2] = \mathbb{E}[|u_{ia}|^2 (\sum_{b=1}^{n} |u_{ib}|^2)] = \E[|u_{ia}|^4] + (n-1)\E[|u_{ib}|^2 |u_{ia}|^2]
\]
we obtain, for $a \neq b$
\[
\E[|u_{ia}|^2 |u_{ib}|^2] = \frac{1}{n(n+1)}
\]


\subsection{Dual Certificates}

With $\cA$ as above, it can be verified that
\[
\frac{1}{m}\E \left[ \cA^* \cA \right] = I - \frac{1}{n+1} I \otimes I = \cS
\]
and we have $\cS^{-1}(X) = X - \frac{1}{n+1}\tr(X)I_n$. Thus, the regular construction of the dual certificate would be\\
\begin{align*}
\frac{1}{m}\cA^* \cA \cS^{-1} (e_1 e_1^*) &= \frac{1}{m} \sum_{i=1}^{m}n(n+1) u_i u_i^* \otimes u_i u_i^* (e_1 e_1^* - \frac{1}{n+1}I_n)\\
&= \frac{n(n+1)}{m}\sum_{i=1}^{m} (|u_{i1}|^2 - \frac{1}{n+1})u_i u_i^*\\
&= \frac{n}{m} \sum_{i=1}^{m} ((n+1)|u_{i1}|^2 -1)u_i u_i^*
\end{align*}

Let $\psi_n = \E\left[ (n)(n+1) ( |u_{i1}|\wedge\frac{3}{\sqrt{n+1}} ) ^4 \right]$. $\psi_n$ is slightly less than 2. Using a construction similar to that found in $\left[5\right]$, we could then take the enhanced certificate to be
\[
Y =  \frac{1}{m} \sum_{i=1}^{m} (2n(n+1)(|u_{i1}|\wedge\frac{3}{\sqrt{n+1}} )^2  - n(2\psi_n -1))u_i u_i^*
\]
We have then the expected value of this sum is 1 in the upper left corner, near to -1 on the rest of the diagonal and zero elsewhere. Furthermore, the contribution of the $|u_{i1}|$ term is capped to not be too large. We thus hope to acquire the same properties of the enhanced dual certificate as in the gaussian case. 

\subsection{Behavior of $Y_T$}

Here we control the quantity $\|Y_T - e_1e_1^*\|_F$. We can re-write the certificate as 
\[
Y =\frac{1}{r} \sum_{k=1}^{r} \sum_{i=1}^{n} (2(n+1)(|u_{i1}^{(k)}|\wedge\frac{3}{\sqrt{n+1}} )^2  - (2\psi_n -1))u_i^{(k)} {u_{i}^{(k)}}^*
\]
where $\{u_i^{(k)}\}_{i=1}^n$ are (indexed by k)  iid Haar distributed on $\mathbb{U}_n$. To show that $\|Y_{T}-e_1 e_1^*\|_{F}$ is small, it is enough to show that
\[
\|\frac{1}{r} \sum_{k=1}^{r} x_k - e_1\|_2
\]
is small, where
\[
x_k =^d \sum_{i=1}^n(2(n+1)(|u_{i1}|\wedge\frac{3}{\sqrt{n+1}} )^2  - (2\psi_n -1)) \bar{u}_{i1} u_i
\]
We have 
\begin{align*}
\E\left[ \|x_k\|^2\right] &= \E\left[ \sum_{i=1}^{n}\left| (2(n+1)(|u_{i1}|\wedge\frac{3}{\sqrt{n+1}} )^2  - (2\psi_n -1))\right|^2 |u_{i1}|^2 \right]\\
&= n\E\left[ \left(4(n+1)^2\left(|u_{i1}|\wedge \frac{3}{\sqrt{n+1}} \right)^4\right)|u_{i1}|^2\right] +\\
& n\E\left[ \left((2\psi_n -1)^2 - 4(n+1)(2\psi_n-1) \left(|u_{i1}|\wedge \frac{3}{\sqrt{n+1}} \right)^2\right)|u_{i1}|^2 \right]\\
&= 4n(n+1)^2\E\left[\left(|u_{i1}|\wedge \frac{3}{\sqrt{n+1}} \right)^4|u_{i1}|^2  \right] + (2\psi_n-1)^2\\
& - 4n(n+1)(2\psi_n-1)\E\left[\left(|u_{i1}|\wedge \frac{3}{\sqrt{n+1}} \right)^2|u_{i1}|^2  \right] \\
&\leq 4n(n+1)^2 \E\left[ |u_{i1}|^6 \right] + (2\psi_n-1)^2 - 4n(n+1)(2\psi_n-1)\E\left[ \left( |u_{i1}|\wedge \frac{3}{\sqrt{n+1}} \right)^4 \right]\\
&=4n(n+1)^2\frac{3!}{n(n+1)(n+2)} + 4\psi_n^2 - 4\psi_n + 1 - 4(2\psi_n-1)\psi_n\\
&= 24\frac{n+1}{n+2} + 1-4\psi_n^2 \leq 24
\end{align*}

Furthermore, we have
\[
\|x_k\|_2 = \left( \sum_{i=1}^n |(2(n+1)(|u_{i1}|\wedge\frac{3}{\sqrt{n+1}} )^2  - (2\psi_n -1)) \bar{u}_{i1}|^2  \right)^{1/2} \leq \sqrt{21}
\]
These facts allow us to apply the vector Bernstein inequality (Theorem 3.5.3) to get that $\|Y_{T}-e_1 e_1^*\|_{F}$ is as small as necessary with probability at least $1- e^{-cr}$ for some constant c. \\

\subsection{Behavior of $Y_{\Tp}$}

We would like to show that $Y_{\Tp} \prec 0$ whp. It is enough to consider $\sup \{\left<x,Y_{\Tp}x\right>; x \in \mathbb{C}\mathbb{S}^n, x_1 = 0 \}$ and we aim to control this quantity via a covering argument. Using rotational invariance, we have
\begin{align*}
\left<x,Y_{\Tp}x\right> =^d \left<e_2,Y_{\Tp}e_2\right> &= \frac{1}{r} \sum_{k=1}^{r} \sum_{i=1}^{n} (2(n+1)(|u_{i1}^{(k)}|\wedge\frac{3}{\sqrt{n+1}} )^2  - (2\psi_n -1))|u_{i2}^{(k)}|^2 \\
&= \frac{1}{r} \sum_{k=1}^{r} \sum_{i=1}^{n} (2(n+1)(|u_{i1}^{(k)}|\wedge\frac{3}{\sqrt{n+1}} )^2)|u_{i2}^{(k)}|^2 - (2\psi_n -1) 
\end{align*}
A straightforward application of Talagrand's inequality fails here. Bernstein's inequality for weakly dependent variables also fails \cite{Three}, so we will use an approach that involves conditioning and Talagrand's inequality. It suffices to show that
\[
\frac{1}{r} \sum_{k=1}^{r} \sum_{i=1}^{n} (2(n+1)(|u_{i1}^{(k)}|\wedge\frac{3}{\sqrt{n+1}} )^2-\phi_{n})|u_{i2}^{(k)}|^2
\]
concentrates well about 0, where 
\begin{align*}
\phi_n &= \E\left[ \frac{1}{r} \sum_{k=1}^{r} \sum_{i=1}^{n} (2(n+1)(|u_{i1}^{(k)}|\wedge\frac{3}{\sqrt{n+1}} )^2)|u_{i2}^{(k)}|^2\right] \\
&= \E\left[ 2n(n+1)(|u_{i1}|\wedge\frac{3}{\sqrt{n+1}} )^2)|u_{i2}|^2 \right] \leq 2
\end{align*}
we have,
\begin{align*}
& \frac{1}{r} \sum_{k=1}^{r} \sum_{i=1}^{n} (2(n+1)(|u_{i1}^{(k)}|\wedge\frac{3}{\sqrt{n+1}} )^2-\phi_{n})|u_{i2}^{(k)}|^2 \\
& =^d \frac{1}{r}\sum_{k=1}^r G(\zeta^{(k)},z^{(k)})\\
&= \frac{1}{r}\sum_{k=1}^{r}\sum_{i=1}^{n} (2(n+1)(|v(z^{(k)})_i|\wedge\frac{3}{\sqrt{n+1}} )^2-\phi_{n})|v(t(v(\zeta^{(k)}),v(z^{(k)})))_i|^2 
\end{align*}
and as before, we consider the surrogate function
\[
\frac{1}{r}\sum_{k=1}^r \tilde{G}(\zeta^{(k)},z^{(k)}) = \frac{1}{r}\sum_{k=1}^{r}\sum_{i=1}^{n} (2(n+1)(|\tilde{v}_1(z^{(k)})_i|\wedge\frac{3}{\sqrt{n+1}} )^2-\phi_{n})|\tilde{v}_2(t(\tilde{v}_1(\zeta^{(k)}),\tilde{v}_1(z^{(k)})))_i|^2 
\]
Now,
\begin{align*}
& \Prob \left( |\frac{1}{r}\sum_{k=1}^r \tilde{G}(\zeta^{(k)},z^{(k)})|\geq t \right) \\
&= \E \left[ \E \left[ \chi_{\{|\frac{1}{r}\sum_{k=1}^r \tilde{G}(\zeta^{(k)},z^{(k)})|\geq t \} } \middle| (z^{(1)},\ldots, z^{(r)} ) \right] \right]\\
&= \E_{z}\left[  \Prob_{\zeta}\left( \left|\frac{1}{r}\sum_{k=1}^r \tilde{G}(\zeta^{(k)}, z^{(k)} )\right|\geq t \right) \right]\\
&\leq \E_{z} \left[ \Prob_{\zeta} \left( \left| \frac{1}{r}\sum_{k=1}^r \tilde{G}(\zeta^{(k)}, z^{(k)} ) - \E_{\zeta} \left[ \frac{1}{r}\sum_{k=1}^r \tilde{G}(\zeta^{(k)}, z^{(k)} ) \right] \right| \geq t - \left|\E_{\zeta} \left[\frac{1}{r}\sum_{k=1}^r \tilde{G}(\zeta^{(k)}, z^{(k)} ) \right]\right| \right) \right]\\
&\leq \E_{z} \left[ \Prob_{\zeta} \left( \left| \frac{1}{r}\sum_{k=1}^r \tilde{G}(\zeta^{(k)}, z^{(k)} ) - f(\{z^{(i)}\}_{i=1}^{r})\right| \geq t - t_1  \right) \chi_{\{|f(\{z^{(i)}\}_{i=1}^{r})| \leq t_1\} } \right]\\
&+ \Prob(|f(\{z^{(i)}\}_{i=1}^{r})| > t_1)\\
\end{align*}
where $f(\{z^{(i)}\}_{i=1}^{r})=\E_{\zeta}\left[\frac{1}{r}\sum_{k=1}^{r}\tilde{G}(\zeta^{(k)},z^{(k)})\right]$.\\

It now suffices to analyze the quantities $\lip_{\zeta}(\tilde{G}(\zeta, z ))$ and $ \E_{\zeta}\left[\tilde{G}(\zeta, z )  \right]$ as functions of $z$.\\
 For $x \in \mathbb{R}^n$ or $\mathbb{C}^{n}$, $g(x) = \sum_{i=1}^{n}a_i |x_i|^2$ and $\|x_1\|_2 + \|x_2\|_2 \leq 2$, we have 
\[
|g(x_1)-g(x_2)| \leq 2\|a\|_{\infty}\|x_1-x_2\|_2
\]
Letting $a_i =  \left(2(n+1)(|\tilde{v}_1(z)_i|\wedge\frac{3}{\sqrt{n+1}} )^2-\phi_{n}\right)$ and noting $\|a\|_{\infty} \leq 20$
\begin{align*}
\left| \tilde{G}(\zeta_1, z ) - \tilde{G}(\zeta_2, z ) \right| &= \left| g(\tilde{v}_2(t(\tilde{v}_1(\zeta_1),\tilde{v}_1(z)))) - g(\tilde{v}_2(t(\tilde{v}_1(\zeta_2),\tilde{v}_1(z)))) \right|\\
&\leq 2 \|a\|_{\infty} \| \tilde{v}_2(t(\tilde{v}_1(\zeta_1),\tilde{v}_1(z))) -  \tilde{v}_2(t(\tilde{v}_1(\zeta_2),\tilde{v}_1(z))) \|_2\\
&\leq 40\lip(\tilde{v}_2)\lip(t|_{\mathcal{B}(0,1)^2})\lip(\tilde{v}_1)\|\zeta_1 - \zeta_2\|_2
\end{align*}
In conclusion
\[
\lip_{\zeta}(\tilde{G}(\zeta, z )) \leq 8*40\frac{\sqrt{\frac{c_1}{c_2}}}{\sqrt{n}}
\]
uniformly in z and thus
\[
\lip(\frac{1}{r}\sum_{k=1}^{r}\tilde{G}(\zeta^{(k)}, z^{(k)} )) \leq \frac{1}{\sqrt{r}}8*40\frac{\sqrt{\frac{c_1}{c_2}}}{\sqrt{n}}
\]
uniformly in $\left(z^{(1)},\ldots,z^{(r)}\right)$. This gives that
\[
\Prob_{\zeta} \left( \left| \frac{1}{r}\sum_{k=1}^r \tilde{G}(\zeta^{(k)}, z^{(k)} ) - f(z^{(1)},\ldots,z^{(r)}) \right| \geq t )  \right) \leq e^{-crnt^2}
\]
for a constant c which depends on $c_i$ but does not depend on z. Now we need to show that $f(z^{(1)},\ldots,z^{(r)})$ concentrates well about its mean and that this mean is very small. We have
\begin{align*}
&f(z^{(1)},\ldots,z^{(r)})\\
& = \frac{1}{r}\sum_{k=1}^{r}\sum_{i=1}^{n}\E_{\zeta} \left[|\tilde{v}_2(t(\tilde{v}_1(\zeta^{(k)}),\tilde{v}_1(z^{(k)})))_i|^2\right]\left(2(n+1)(|\tilde{v}_{1}(z^{(k)})_i |\wedge \frac{3}{\sqrt{n+1}})^2 - \phi_n \right)
\end{align*}
Let  
\[
h(z) = \{\E_{\zeta} \left[|\tilde{v}_2(t(\tilde{v}_1(\zeta),\tilde{v}_1(z)))_i|^2\right]\}_{i=1}^{n}.
\]
and
\[
p(z) = \{(2(n+1)(|\tilde{v}_{1}(z)_i |\wedge \frac{3}{\sqrt{n+1}})^2 - \phi_n )\}_{i=1}^{n}
\]
First, using the following facts,
\begin{align*}
&\E\left[ \tilde{v}_1(\zeta)_i \right] = 0\\
&\E\left[ \tilde{v}_1(\zeta)_a \tilde{v}_1(\zeta)_b \right] = 0, a \neq b\\
&\E\left[|\tilde{v}_1(\zeta)_i |^2\right] \leq \frac{1}{n}\\
&\E\left[ |\left< \tilde{v}_1( \zeta),y \right>|^2 \right] \leq \|y\|_2^2 \frac{1}{n}\\
& \E\left[ 2\tilde{v}_1(\zeta)_i \bar{y}_i \left<\tilde{v}_1(\zeta),y\right> \right] =0  
\end{align*}
for any $y \in \mathbb{C}^n$, we establish
\begin{align*}
&\E_{\zeta} \left[|\tilde{v}_2(t(\tilde{v}_1(\zeta),\tilde{v}_1(z)))_i|^2\right] \\
&\leq \E_{\zeta} \left[\frac{1}{c_2}|(t(\tilde{v}_1(\zeta),\tilde{v}_1(z)))_i|^2\right]\\
&\leq \frac{1}{c_2} \E_{\zeta}\left[  |\tilde{v}_1(\zeta)_i|^2 + |\tilde{v}_1(z)_i|^2 \left|\left< \tilde{v}_1(z),\tilde{v}_1(\zeta) \right> \right| ^2 - 2\Re(\tilde{v}_1(\zeta)_i \bar{\tilde{v}}_1 (z)_i  \left< \tilde{v}_1(\zeta),\tilde{v}_1(z) \right>) \right]\\
& \leq \frac{1}{c_2}\left[ \frac{1}{n} + |\tilde{v}_1(z)_i|^2 \|\tilde{v}_1(z)\|_2^2 \frac{1}{n} \right] \leq \frac{2}{c_2 n}
\end{align*}
Thus, for any z, $\|h(z)\|_{\infty} \leq \frac{2}{c_2 n}$.\\

Now we shall compute $\lip(\sum_{i=1}^{n}h_i(z)p_i(z))$ directly:
\begin{align*}
&\left| \sum_{i=1}^{n}h_i(z_1)p_i(z_1) - \sum_{i=1}^{n}h_i(z_2)p_i(z_2) \right|\\
&\leq \|h(z_1)\|_{\infty}\sum_{i=1}^{n}\left|  p_i(z_1)-p_i(z_2) \right| + \|p(z_2)\|_{\infty}\sum_{i=1}^{n}\left| h(z_2)-h(z_2) \right|\\
&\leq 2(n+1)\|h(z_1)\|_{\infty} \sum_{i=1}^{n}\left|( |\tilde{v}_1(z_1)_i|\wedge \frac{3}{\sqrt{n+1}})^2 - (|\tilde{v}_1(z_2)_i|\wedge \frac{3}{\sqrt{n+1}})^2\right|+\\
& \|p(z_2)\|_{\infty} \E_\zeta\left[ \sum_{i=1}^{n}\left| |\tilde{v}_2(t(\tilde{v}_1(\zeta),\tilde{v}_1(z_1)))_i|^2 - |\tilde{v}_2(t(\tilde{v}_1(\zeta),\tilde{v}_1(z_2)))_i|^2  \right|  \right]\\
&\leq 2(n+1)\|h(z_1)\|_{\infty}2(\||\tilde{v}_1(z_1)|\wedge \frac{3}{\sqrt{n+1}} - |\tilde{v}_1(z_2)|\wedge \frac{3}{\sqrt{n+1}}  \|_2)+\\
&\|p(z_2)\|_{\infty} \E_{\zeta}\left[  2\| \tilde{v}_2(t(\tilde{v}_1(\zeta),\tilde{v}_1(z_1)))-\tilde{v}_2(t(\tilde{v}_1(\zeta),\tilde{v}_1(z_2)))  \|_2 \right]\\
&\leq \left[ 2(n+1)\|h(z_1)\|_{\infty}2\lip(\tilde{v}_1) + \|p(z_2)\|_{\infty} 2 \lip(\tilde{v}_2)\lip(t|_{\mathcal{B}(0,1)^2})\lip(\tilde{v}_1) \right]\|z_1-z_2\|_2\\
&\leq \left(16\frac{n+1}{ n}\frac{\sqrt{c_1}/c_2}{\sqrt{n}} + 320 \frac{\sqrt{\frac{c_1}{c_2}}}{\sqrt{n}}\right)\|z_1-z_2\|_2
\end{align*}
Thus,
\[
\lip(f(z^{(1)},\ldots,z^{(r)})) =  \text{O}\left( \frac{\sqrt{c_1}}{c_2} \frac{1}{\sqrt{rn}} \right)
\]
This will allow us to get the desired concentration of $f(z^{(1)},\ldots,z^{(r)})$ around its mean via Talagrand's inequality. Namely, we obtain
\[
\Prob\left( \left| f(z^{(1)},\ldots,z^{(r)}) - \E\left[ f(z^{(1)},\ldots,z^{(r)}) \right] \right| \geq t \right) \leq e^{-crnt^2} 
\]
for a constant c which depends on $c_i$.\\

Let $F = \frac{1}{r}\sum_{k=1}^r G(\zeta^{(k)},z^{(k)})$ and  $\tilde{F} = \frac{1}{r}\sum_{k=1}^r \tilde{G}(\zeta^{(k)},z^{(k)})$. Then $ \E\left[ f(z^{(1)},\ldots,z^{(r)}) \right] = \E\left[ \tilde{F} \right]$ and note $\E\left[F \right] = 0$. Since both $F$ and $\tilde{F}$ are bounded and differ on a set of exponentially small probability, for any valid choice of $c_i$, $\lim_{n \rightarrow \infty} \E\left[\tilde{F}\right] = 0$ and so having fixed t apriori, for n large enough
\[
\Prob\left( \left| f(z^{(1)},\ldots,z^{(r)})\right| \geq \frac{t}{2} \right) \leq e^{-crn(t/4)^2} 
\]

%
Taking $t_1 = \frac{t}{2}$, this implies 
\[
\Prob \left( |\frac{1}{r}\sum_{k=1}^r \tilde{G}(\zeta^{(k)},z^{(k)})|\geq t \right)  \leq e^{-crn(t-\frac{t}{2})^2} + e^{-crn(t/4)^2 }
\]
Now using that $F$ and $\tilde{F}$ differ on a set of  probability at most $\text{O}(re^{-\gamma n})$, we have
\begin{align*}
\Prob \left( |F|\geq t \right)  &\leq \Prob\{ |\tilde{F}| \geq t - |\tilde{F}-F|\}\\
&\leq \Prob\{F \neq \tilde{F}\} +  \Prob\{ |\tilde{F}| \geq t \}\\
&\leq  \text{O}(re^{-\gamma n}) + e^{-crn(t/2)^2} + e^{-crn(t/4)^2 }
\end{align*}
Therefore, we have established
\[
\mathbb{P}\{ \left<x,Y_{\Tp}x \right> \geq t + \phi_n - (2\psi_n-1) \} \leq \text{O}(re^{-\gamma n}) + e^{-crn(t/2)^2} + e^{-crn(t/4)^2 }
\]
To get an arbitrarily fast exponential rate of concentration, fix $\gamma$ to be as large as needed by choosing $c_i$ appropriately, then fix r large enough. Note that $\phi_n \leq 2$ and $\psi_n$ is very close to 2 so that $\phi_n - (2\psi_n-1) \approx -1$. Choosing an appropriate t, we get that $Y_{\Tp}$ is negative definite with high probability via the standard covering argument, which completes the proof of the main theorem. 

\section{Discussion}
We show that pure states can be recovered from few full-rank observables using the PhaseLift algorithm. We note that these results can be extended to show noise stability by using a modified convex program, as well as uniformity over signals $x \in \mathbb{C}^n$. Finally, a very similar proof would yield that rank-$k$ states $X_k = \sum_{i=1}^k \lambda_i x_i x_i^*$ can be recovered from $m = O(kn)$ measurements, by establishing an RIP-1 property for rank $2k$ matrices and a using a dual certificate motivated by $\cA^* \cA(X_k)$.

\section{Acknowledgements}
We acknowledge fruitful discussions with Emmanuel Candes and Xiaodong Li. 

\bibliography{ucbtest.bib}
\bibliographystyle{plain}

\end{document}